\documentclass{jkas0}
\usepackage{amsmath}

\def\beginpage{207} 
\def\received{July 3, 2015} 
\def\accepted{July 31, 2015} 
\date{Received \received ; accepted \accepted}

\newcommand\ion[2]{{#1}\,{\sc #2}} 

\def\farcm{\hbox{$.\mkern-4mu^\prime$}}
\def\farcs{\hbox{$.\!\!^{\prime\prime}$}}

\title{
Lee Sang Gak Telescope (LSGT):\\ A Remotely Operated Robotic Telescope for
Education and Research at Seoul National University
}

\author[1,2,3]{Myungshin~Im}
\author[1,2]{Changsu~Choi}
\author[2]{Kihyun~Kim}

\affil[1]{Center for the Exploration of the Origin of the Universe, Department of Physics and Astronomy, Seoul National University, Gwanak-gu, Seoul 151-742, Korea; \email{mim@astro.snu.ac.kr}}
\affil[2]{Astronomy Program, Department of Physics and Astronomy, Seoul National University, Gwanak-gu, Seoul 151-742, Korea}
\affil[3]{Korea Institute for Advanced Study, 85 Hoegiro, Dongdaemun-gu, Seoul 130-722, Republic of Korea}

\def\corrauthor{
M. Im
}

\def\runningauthor{
Im et al.
}

\def\runningtitle{
Lee Sang Gak Telescope: A Robotic Telescope for Education and Research
}

\def\keywords{
telescopes --- instrumentation: detectors --- methods: observational --- technique: photometric
}

\def\abstracttext{
  We introduce the Lee Sang Gak Telescope (LSGT), a remotely operated, robotic 0.43-meter telescope.
  The telescope was installed at the Siding Spring Observatory, Australia,
 in 2014 October, to secure
 regular and exclusive access to the dark sky and excellent atmospheric
 conditions
 in the southern hemisphere from the Seoul National University (SNU) campus.
  Here, we describe the LSGT system and its performance, present example images
 from early observations, and discuss a future plan to upgrade the system.
  The use of the telescope includes (i) long-term monitoring observations of nearby galaxies, active galactic nuclei, and supernovae; (ii) rapid follow-up observations of
 transients such as gamma-ray bursts and gravitational wave sources; and (iii) observations
 for educational activities at SNU. Based on observations performed so far,
 we find that the telescope is capable of providing images to a depth of $R=21.5$ mag
 (point source detection) at 5-$\sigma$ with 15 min total integration time under  good observing conditions.
}

\begin{document}
\jkashead 


\section{Introduction\label{sec:intro}}

  To obtain high quality data for astronomical research,
  it is necessary to install telescopes at locations offering dark sky,
 low humidity, and stable atmospheric
 conditions. Such places are usually found in remote locations,
  on mountain-tops
 that do not allow easy access, forcing astronomers to make long trips to
 these locations and making it difficult to carry out observations flexibly
 such as a rapid observation of new interesting astronomical sources.
  However, recent advances in computer networking and
 remote control technologies have made it possible to operate telescopes
 remotely from places such as cities where most astronomers
 reside.
  Nowadays, professional astronomers often carry out
 observations from a remote location (e.g., their office). Amateur
 astronomers are also setting up remotely operated telescopes in locations favorable for
 astronomical observations.

   Seoul National University (SNU) has several optical telescopes on campus that
  are equipped with optical imagers and a low resolution spectrograph.
   These telescopes have been used for educational activities of SNU students
 and occasionally for research projects. However, the use of these telescopes
 has suffered greatly from light-pollution due the
 fast development of the metropolitan Seoul area in the last several decades.
  The sky brightness at SNU has been reported as
  $B \simeq 17$ -- $18$ mag arcsec$^{-2}$ and $V \simeq 16$ -- 17.3 mag arcsec$^{-2}$
  (Lee et al. 2009), several hundred times brighter than the night sky of
  astronomical observatories at remote locations
  (e.g., Leinert \& Mattila 1998; Patat 2003; Pedani 2009; Aceituno et al. 2011).
  The bright night sky limits observations
 at the SNU campus to only bright objects with $R \lesssim 18.3$ mag with
  a 0.6-meter telescope (Choi et al. 2014; 5-$\sigma$ detection limit of a point source
  at 15 min total integration).

   In order to access dark skies with excellent astro-climate, we have installed
  instruments on telescopes at several remote sites. Examples include
  the Seoul National University 4k$\times$4k Camera
  (SNUCAM) on the 1.5-meter telescope of the Maidanak observatory, Uzbekistan (Im et al.
  2010), and the Camera for QUasars in EArly uNiverse (CQUEAN) of the 2.1-meter telescope at
  the McDonald observatory (Park et al. 2010; Kim et al. 2010; Lim et al. 2013).
   However, the use of these facilities
  has been rather limited, mostly due to high demands for the telescopes for other
  research projects.
   In order to gain regular, exclusive access of the southern hemisphere
  dark sky, we recently installed a remotely operated, robotic
  telescope at the Siding Spring Observatory (SSO) in Australia.
  We named this telescope as ``Lee Sang Gak Telescope" (LSGT)
  to honor the recently retired
  SNU professor who donated a significant portion of the funding
  that made this telescope possible. LSGT is now regularly used for
  observational classes and research projects at SNU.

   LSGT is anticipated to be extensively used by both astronomers and students.
   To help the potential telescope users to plan observations with LSGT,
   we describe in this paper the overall characteristics of LSGT,
  its performance, and highlights from its usage so far.

\begin{figure}[t!]
\centering
\includegraphics[width=83mm]{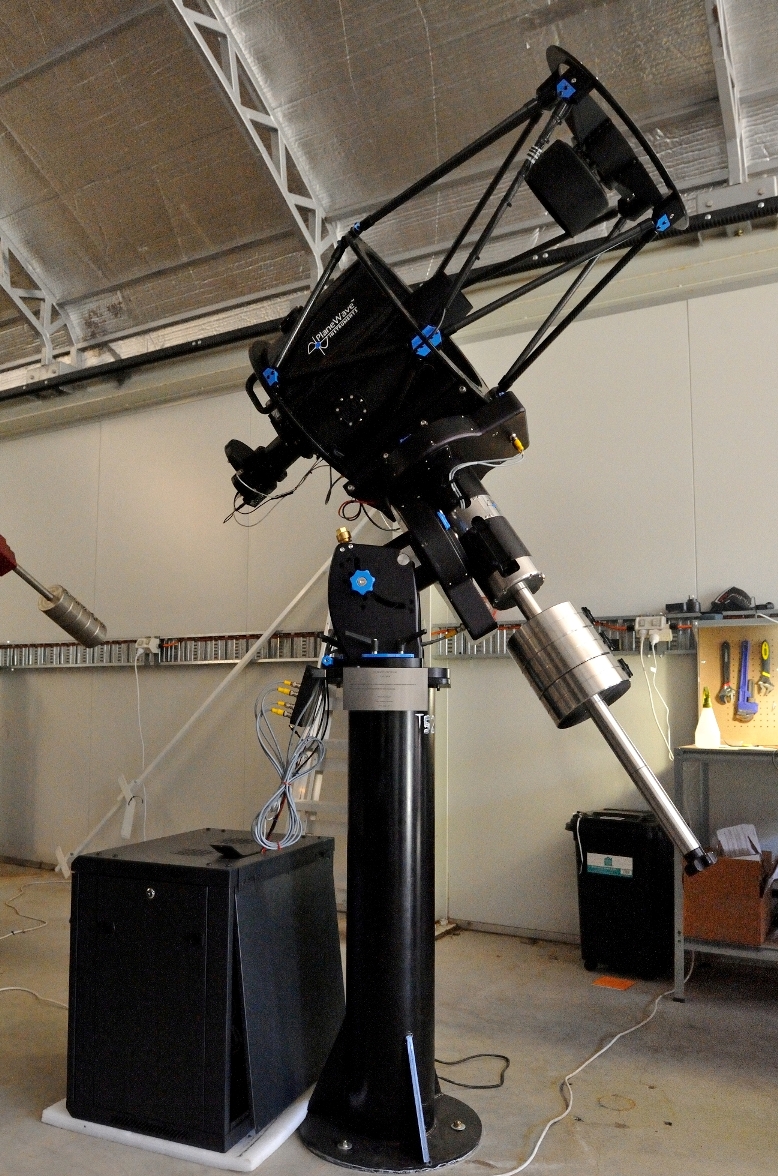}
\caption{The Lee Sang Gak Telescope, shown shortly after its installation at the
Siding Spring Observatory, Australia in 2014 October. }
\end{figure}

\section{System}

  LSGT (Figure 1) is a Corrected Dall-Kirkham (CDK) design
 telescope with a 0.43-meter (17-inch) diameter primary mirror. It
 is manufactured by the PlaneWave Instruments\footnote{\url{http://planewave.com}}
 and matched with a PlaneWave Instrument's Ascension 200HR mount. The CDK design adopts an ellipsoidal primary mirror,
 a spherical secondary mirror, and a combination of two lenses
 to produce a coma-free
 flat field with no off-axis astigmatism over a 52mm diameter circle. Given the good
  performance of the mount and a need to rapidly observe targets,
  we opted not to use an auto-guiding system.
   The effective focal ratio of the telescope is f/6.8, and the telescope is designed to achieve a good, distortion-free image quality over
    1 degree diameter circle.
  Figure 2 shows the combined throughput of
 the reflectivity of the mirrors and the transmittance of the lenses. The data
 are kindly provided by PlaneWave Instruments.

     The currently available instrument is the ST-10XME camera of
 the SBIG Astronomical Instruments (a division of Diffraction Limited). This camera uses the KAF-3200ME chip, which
 offers a peak quantum efficiency (QE) of 85\%
 and a pixel size of 6.8 $\mu$m in a 2184 x 1472 layout.
 On LSGT, the pixel scale translates to $0\farcs48$, and a field of view of
 $17\farcm5 \times 11\farcm8$.
  Another camera, Starlight Xpress's SXVR-H36, is also
 available on demand. The CCD chip, the Truesense's KAI-16000 interline CCD,
 has a dimension of 36.3\,mm $\times$ 24.4\,mm ($4904 \times 3280$ pixels)
 with a physical pixel size of 7.4 $\mu$m. On LSGT, it gives a pixel scale of
 $0\farcs52$ and a field of view of $32\farcm7 \times 26\farcm2$. The QEs
 of these two cameras, along with a planned future upgrade (Section 6), are plotted in Figure 3.

\begin{figure}[t!]
\centering
\includegraphics[trim=2mm 1mm 12mm 2mm, clip, width=83mm]{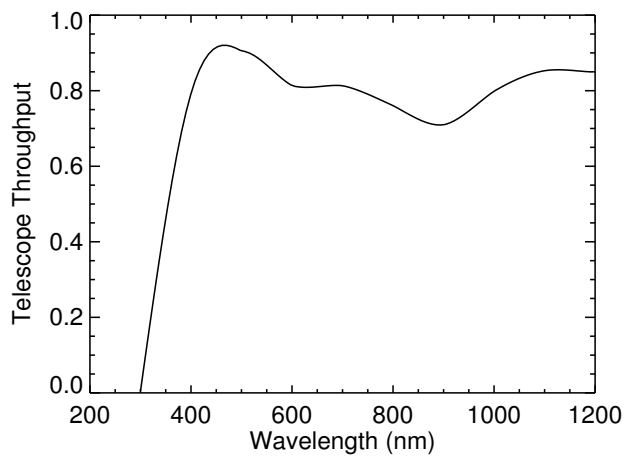}
\caption{The overall throughput of the optical telescope system of LSGT.}
\end{figure}

\begin{figure}[t!]
\centering
\includegraphics[trim=2mm 1mm 12mm 2mm, clip, width=83mm]{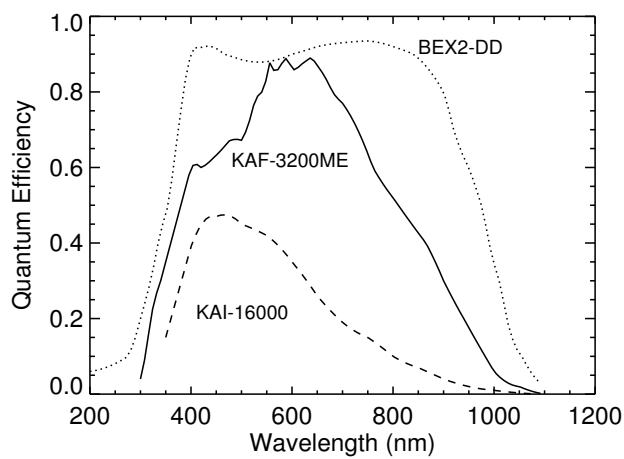}
\caption{QEs of the CCDs on different cameras that have been in use or planned
for LSGT. The current camera uses the KAF-3200ME chip (solid line), while we plan to
upgrade it with a new camera that is equipped with a deep-depletion chip (dotted line). The data are taken from the manufacturer's data sheet.}
\end{figure}

  Table 1 summarizes the characteristics of the cameras on LSGT.
  The standard $UBVRI$ and luminance, H$\alpha$, \ion{S}{ii}, and \ion{O}{iii} filters
  are currently available.
  The telescope is installed in a roll-off-roof dome, together with other telescopes
 that are managed by iTelescope.Net Pty. Ltd.\footnote{\url{http://itelescope.net}}.
  The management of the telescope
 by the private company saves efforts of SNU personnel.

\begin{table*}[!t]
\caption{LSGT Camera Specifications}
\centering
\begin{tabular}{lcccccccc}
\toprule
 Camera  &  Pixel scale &  Field of view  & RN  &  Dark current & Peak QE & Gain   & Full well  & Readout time\\
         &  arcsec      &   arcmin$^{2}$  & e$^{-}$  &   e$^{-}$/pix/sec  &         & e$^{-}$/DN  & e$^{-}$         &   sec       \\
\midrule
 ST-10XME$^{\rm a}$  &  $0.48$  &  $17.5 \times 11.8$ & 10.0$^{\rm b}$  & 0.01 -- 0.5$^{\rm b,c}$   & 0.85  & 1.42$^{\rm b}$    &  $\sim$78,000$^{\rm b}$ & $\sim$20$^{\rm b}$ \\
  SXVR-H36            &  $0.52$  &  $32.7 \times 26.2$ &  9$^{\rm d}$   & 0.01$^{\rm d}$  &  0.45 &  0.4$^{\rm d}$   &  30,000$^{\rm d}$  &  $\sim10$ \\
\bottomrule
\end{tabular}
\tabnote{
 $^{\rm a}$  Operating camera as of 2015 August 31.
\\ $^{\rm b}$ Measurements with CCD temperature at -10$^{\circ}{\rm C}$.
\\ $^{\rm c}$ For most of the pixels, $< 0.05$ e$^{-}$/pix/sec. But, 6 \% are warm pixels with 0.05 - 0.5 e$^{-}$/pix/sec and 2\% with $> 0.5$ e$^{-}$/pix/sec.
\\ $^{\rm d}$ Manufacturer's data.
}
\end{table*}

\begin{figure}[t!]
\centering
\includegraphics[width=83mm]{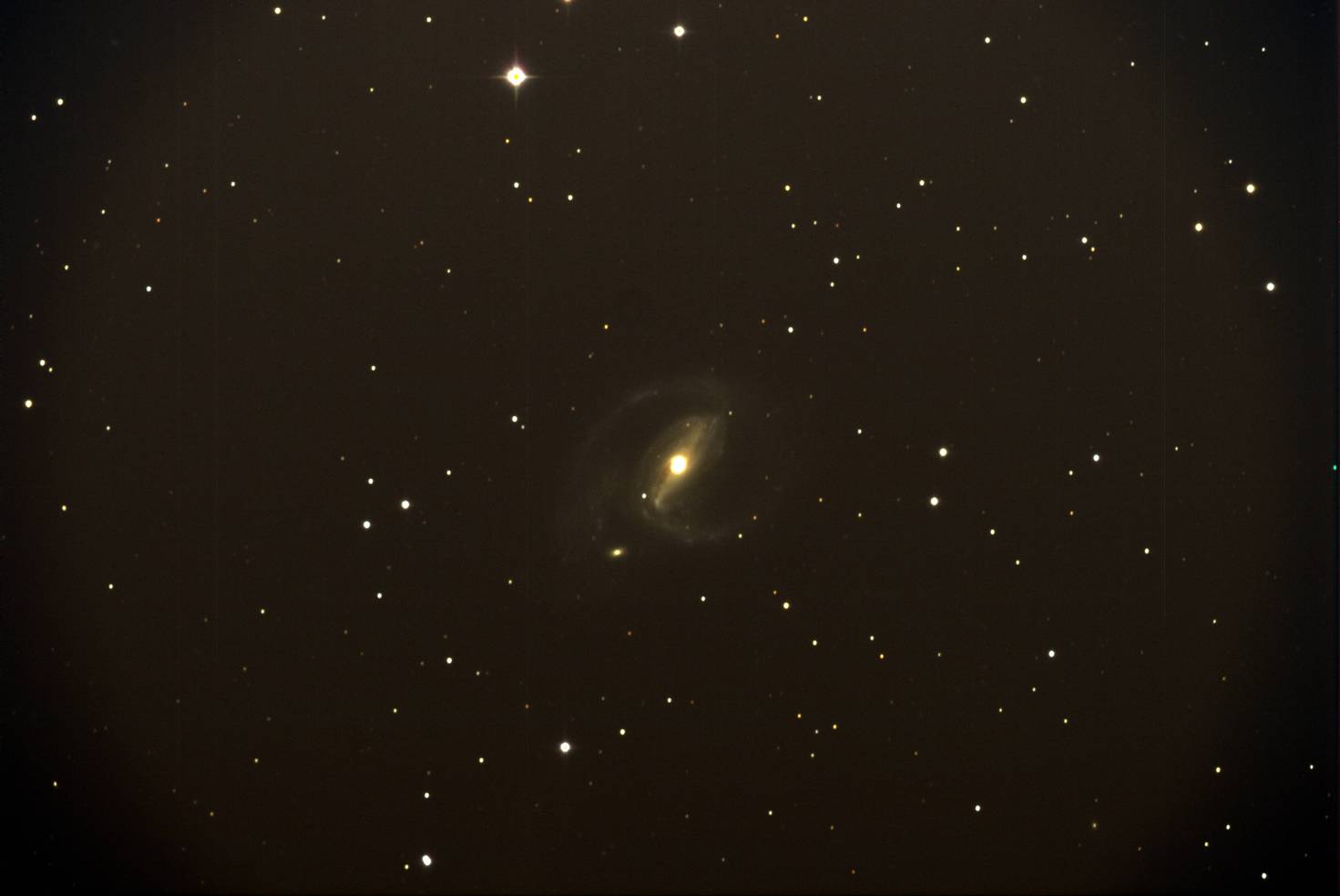}
\caption{A $BVR$ composite color image of NGC 1097 taken with the SXVR-H36 camera.
 While the image quality is good over the whole field, vignetting is visible near the corners
 of the image.}
\end{figure}

\section{Operation and Data Reduction}

 Observations have been carried out either robotically by writing a ``plan" script that lists a number
 of commands that specify the observation of a target, or by typing in
 observational commands remotely in real-time. Under the robotic
 operation, the observations are executed at the time specified in the plan automatically.
 In both
 cases, we access the system through a standard web browser system that is set up by
 iTelescope.Net.

  The focusing of the telescope is done by running an auto-focus program
 over a target field or a field with bright stars near the target. The typical
 overhead for the auto-focusing is about $\sim$3 minutes, including the telescope
 slew time. The command scripts allow us to skip focusing through different
 filters by adopting a pre-defined focus offset, and to specify the frequency of
 the focusing. Typical observations require focusing approximately
 every two hours.
   Bias and dark frames are taken regularly during daytime or cloudy nights
   with the cover of the telescope and
  the dome closed. The flats are taken on demand several times per month using a flat-field screen. The flat-field pattern shows little variation
  day-by-day so the sparse acquisition of the flat field image is adequate.
   As soon as the data are taken, the system performs
 the bias/dark subtraction and flat-fielding with a
 standard set of calibration files. The raw images as well as the calibrated
 images become available soon after the observation (typically a few minutes after)
 for download.

\section{Performance}

 Since the installation of LSGT at SSO in 2014 October, we have been improving
 the system performance. Here, we report the current performance of the telescope  based on observational data.

\subsection{SXVR-H36}

  Early test observations in 2014 October and November were carried out using the SXVR-H36 camera. Figure 4 shows a test image of NGC 1097 taken with this camera.
   $B, V,$ and $R$ band images were combined to create the composite image with
  300 sec exposure per band. Basic calibration
  of bias and dark subtraction and flat-fielding was done.
  The overall quality of the image is good, but a vignetting pattern
  can be seen near the corners of the chip.
   From the image, we measure a 5-$\sigma$ limiting
  magnitude for point sources is $V=20$ mag with 300 sec exposure at
  $3\farcs0$ seeing and a dark night.
   The optical alignment was not optimal during the test observations with the SXVR-H36
   camera, but it allowed a uniform image quality over the camera field of view
   with the $\sim 3\farcs0$ seeing.

\subsection{SBIG ST-10XME and Limiting Magnitude}

  Since {2014 November 14}, LSGT observations have been carried out
 using the ST-10XME camera.
  We decided to use the ST-10XME camera for regular operations
  partly due to the better sensitivity of the camera in comparison to
 the SXVR-H36 camera, and due to the availability of a filter wheel
 that can house ten filters.

 Figure 5 summarizes the seeing values for
  the $R$-band images taken during 2014 November 26 through 2015 April 17.
     The optical
  alignment has been improved during this period of operation, and we sometimes
  achieved seeing conditions of $1\farcs8$.
   With the ST-10XME camera, we achieved
  limiting magnitudes of $R=21.5$ mag for a 15 min exposure,
  with a clear dark night and a seeing size of 2.2 arcsec.

\begin{figure}[t!]
\centering
\includegraphics[trim=3mm 2mm 15mm 3mm, clip, width=83mm]{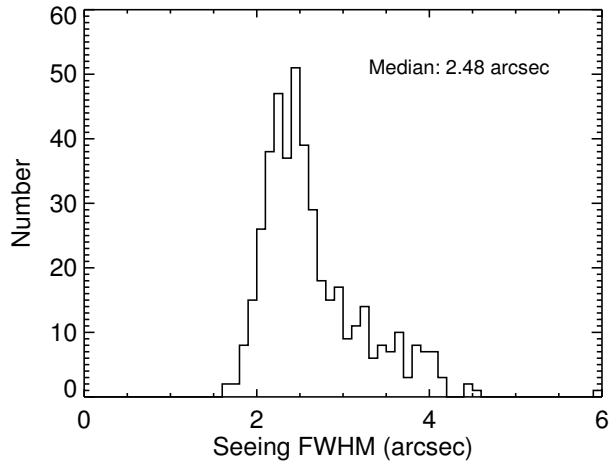}
\caption{The $R$-band seeing statistics
between 2014 November 26 through 2015 April 17. The median seeing is $2\farcs48$.}
\end{figure}

\subsection{Optical Performance and Tracking Accuracy}

  In Figure 6, we show the $R$-band images of the point spread function (PSF)
  at different
 parts of the field of view. The PSFs are constructed using stars in each section of the
 image. An exposure time of 60 sec is used for this image.
  We find that the image quality is
 uniform over the field of view of ST-10XME with a PSF FWHM of  $2\farcs2$.
  However, stars appear slightly elongated
 (an axis ratio, $b/a$, of $\sim$0.95). The PSF elongation is also visible in
 the images taken with much shorter exposures (5 sec). Therefore, we conclude that the elongation is caused mainly by the optics of the telescope.

\begin{figure}[t!]
\centering
\includegraphics[width=82.3mm]{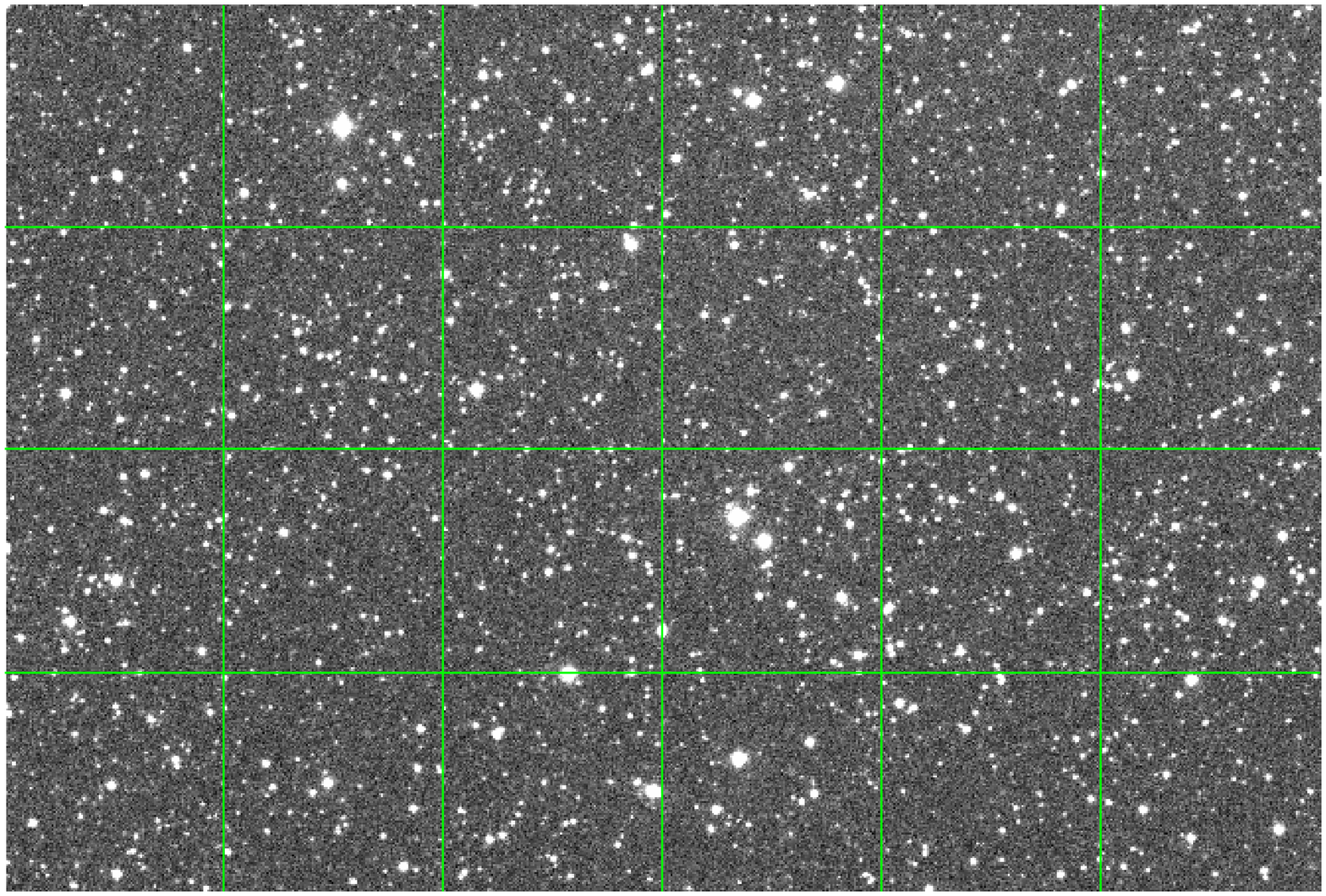}
\includegraphics[width=81mm]{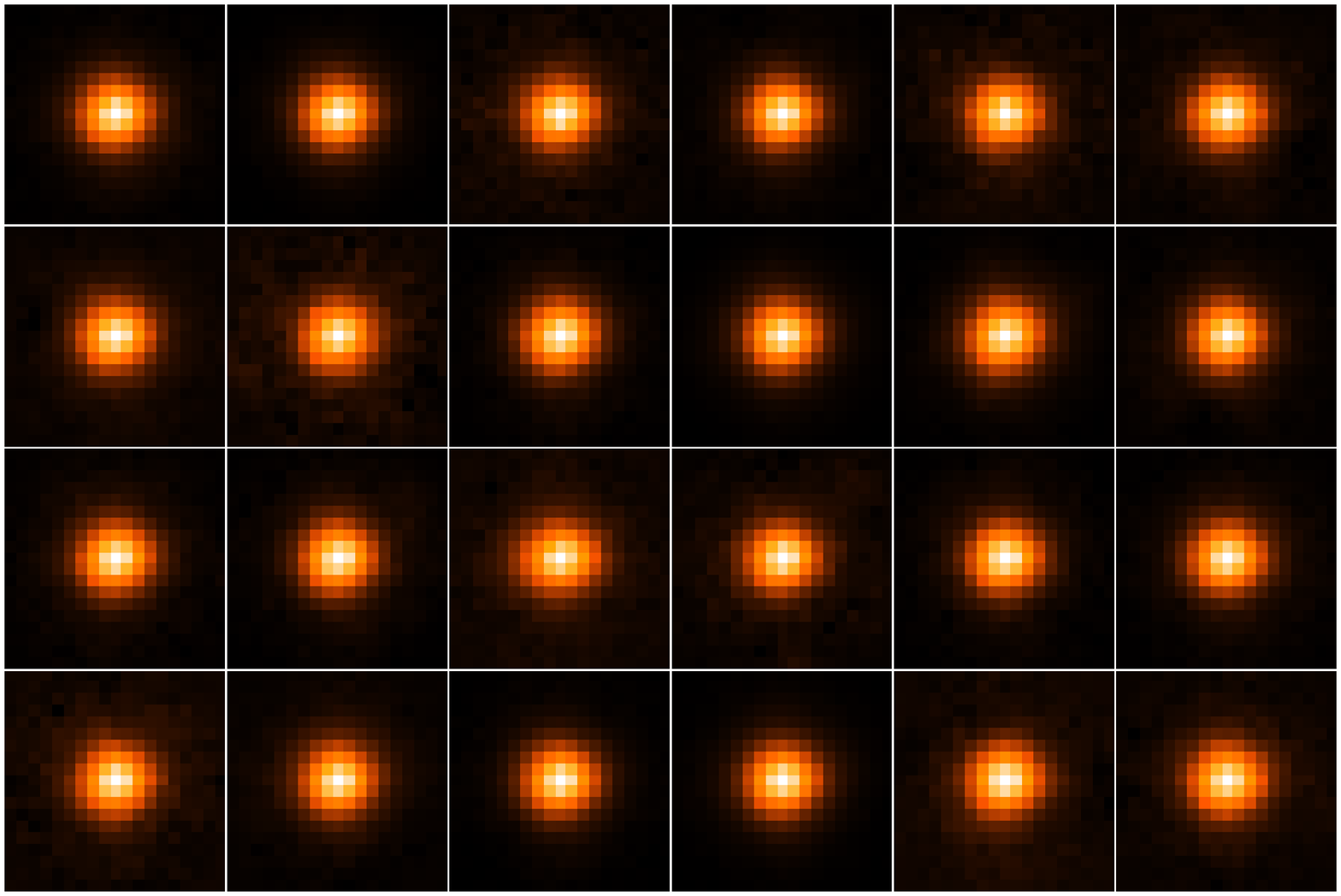}
\caption{The PSF image at different sections of the field of view of the ST-10XME camera. The PSF shape is uniform over the camera field of view, but is slightly elongated, which is partly due to the inaccuracy in tracking, but mostly due to the telescope optics.}
\end{figure}

  Additionally, we note that the PSF can be elongated due to the inaccuracy in
  tracking for long exposure times.
  We examined the tracking accuracy using
 a series of images taken over a long period of time. If the tracking is perfect, we expect to
 find that a star will be located at the same position on the chip in successive frames.
  Otherwise, the star will move toward a certain direction
 in successive exposures.
  We find a drift of about $0\farcs2$/min
 along both the RA and Dec directions, with the direction of the drift changing depending on the location on the sky.
  This test indicates a total drift of $\sim 0\farcs3$/min.

  We calculate how this kind of tracking error elongates the PSF.
  Let us assume that the telescope pointing drifts toward a single
 direction at a rate of $k$
 in units of arcsec/min. For a given exposure time $t_{exp}$, this translates
 into a movement of a star position by an amount of $k \, t_{exp}$ arcmin, which
 causes an elongation of the stellar image in that direction. Assuming that
 the PSF profile, $\mathrm{PSF}(x)$, can be approximated with a Gaussian
 function, the elongated PSF profile can be described as,
\begin{equation}
\begin{split}
 \mathrm{PSF}(x) & \sim \frac{1}{\sqrt{2}\sigma} \int_{-t_{exp}/2}^{t_{exp}/2}
 \, \exp\left[-\frac{(x - k\,t)^{2}}{2 \sigma^{2}}\right] \frac{\mathrm{d}t}{t_{exp}} \\
 \propto\ & \mathrm{erf}\left(\frac{0.5\,k\,t_{exp} - x}{\sqrt{2}\sigma}\right) -
 \mathrm{erf}\left(\frac{-0.5\,k\,t_{exp} - x}{\sqrt{2}\sigma}\right) ,
\end{split}
\end{equation}
where $\sigma=\mathrm{FWHM}/2.35$, and $x$-direction is taken as the  direction of the
elongation. The resulting function resembles the Gaussian form closely, and the FWHM and
$b/a$ values
of the elongated PSF can be derived numerically using Equation (1).

\begin{figure}[t!]
\centering
\includegraphics[trim=2mm 2mm 14mm 4mm, clip, width=83mm]{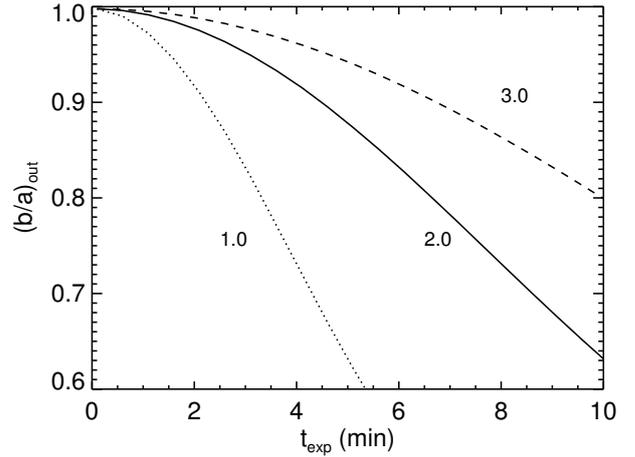}
\caption{The PSF elongation (axis ratio) due to a drift in the telescope pointing
as a function of the exposure time ($t_{exp}$). The numbers next to each line
indicate the FWHM (arcsec) of a round PSF without the pointing drift. The drift rate is assumed to
be $0\farcs3$/min.}
\end{figure}

   In Figure 7, we show how the axis ratio changes a function of $t_{exp}$ according
   to Eq. (1).
    The current system hardly achieves a PSF FWHM better than
 $2\farcs0$.
 Under such conditions, a 180 sec exposure can create an elongated PSF with
 $b/a \simeq 0.95$. For seeing values of FWHM $\sim$ $3\farcs0$,
 the PSF elongation is not noticeable even with an exposure of around
 5 min.
  We conclude that the current system can be used to take
 images with 180 sec exposure, without distorting the PSF shape worse than $b/a = 0.95$.

\section{Science Programs and Current Usage}

  Currently, there is one regular science program running on the telescope.
  It is the Intensive Monitoring Survey of Nearby Galaxies (IMSNG).
  IMSNG performs high cadence imaging observations of nearby galaxies using
  facilities at several different observatories. By using telescopes
  at multiple time zones, it is possible to achieve very high cadence
  monitoring of nearby galaxies with time intervals as short as 2-3 hours.
  Since 2014 December, we have been using LSGT to monitor nearby galaxies at
  distances less than 50 Mpc. These galaxies are chosen to have high
  near-ultraviolet (NUV) fluxes, implying high star formation rates. The scientific
  aim of the project is to detect transients, such as supernovae,
  in a very early phase
  to gain insights about their energetics.
   Through this program, we succeeded in detecting SN 2015F in a very
  early phase at a daily cadence (Im et al. 2015).
   Figure 8 shows NGC 2442 before and after
  the appearance of SN 2015F.

    We have also taken other verification images to evaluate the performance
   of LSGT. In Figure 9, we show single frame
   images of NGC 2467 (a star forming region), NGC 2207 (an interacting galaxy
   pair), and a mosaic image of Omega Centauri (NGC 5139). During SNU observational
   classes for astronomy majors, LSGT has been used to study
   brightness variations of asteroids,
   color-magnitude diagrams of stellar clusters, surface brightness profiles of
   nearby galaxies, and light curves of variable stars and exoplanet transit
   events.

\begin{table}[t!]
\caption{Current LSGT Performance}
\centering
\begin{tabular}{lcc}
\toprule
 Limiting magnitude  & Seeing  &  Tracking accuracy \\
\midrule
 $R=20.5^{a}$ -- $21.5^{b,c}$  &  $2'' - 3''$  &  $0\farcs3$/min   \\
\bottomrule
\end{tabular}
\tabnote{
 $^{\rm a}$  5-$\sigma$ detection of point source with 15 min exposure,  $2\farcs2$ seeing, and full moon.
\\ $^{\rm b}$  Same as above, but with new moon.
\\ $^{\rm c}$  The limiting magnitudes in $B$ and $V$ (under new moon)
are about the same as the $R$-band value.
}
\end{table}

\begin{figure}[t!]
\centering
\includegraphics[width=83mm]{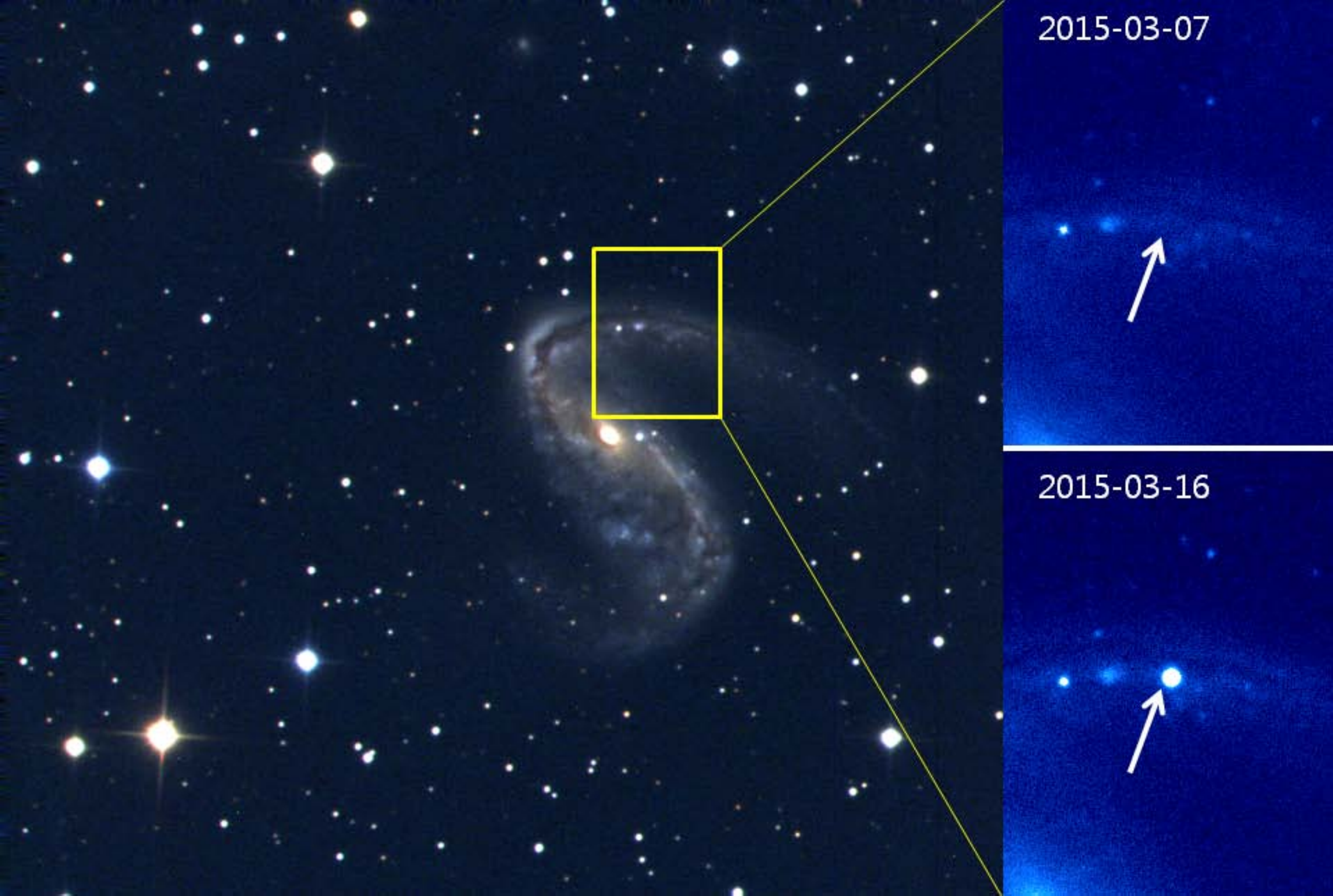}
\caption{BVR composite image of NGC 2442 and the location of SN 2015F
before and after its emergence ($R$-band).}
\end{figure}

\begin{figure}[t!]
\centering
\includegraphics[width=82mm]{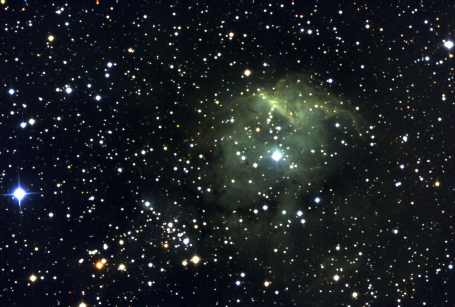} \\
\includegraphics[width=82mm]{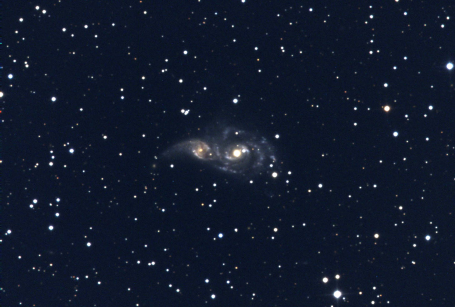} \\
\includegraphics[width=82mm]{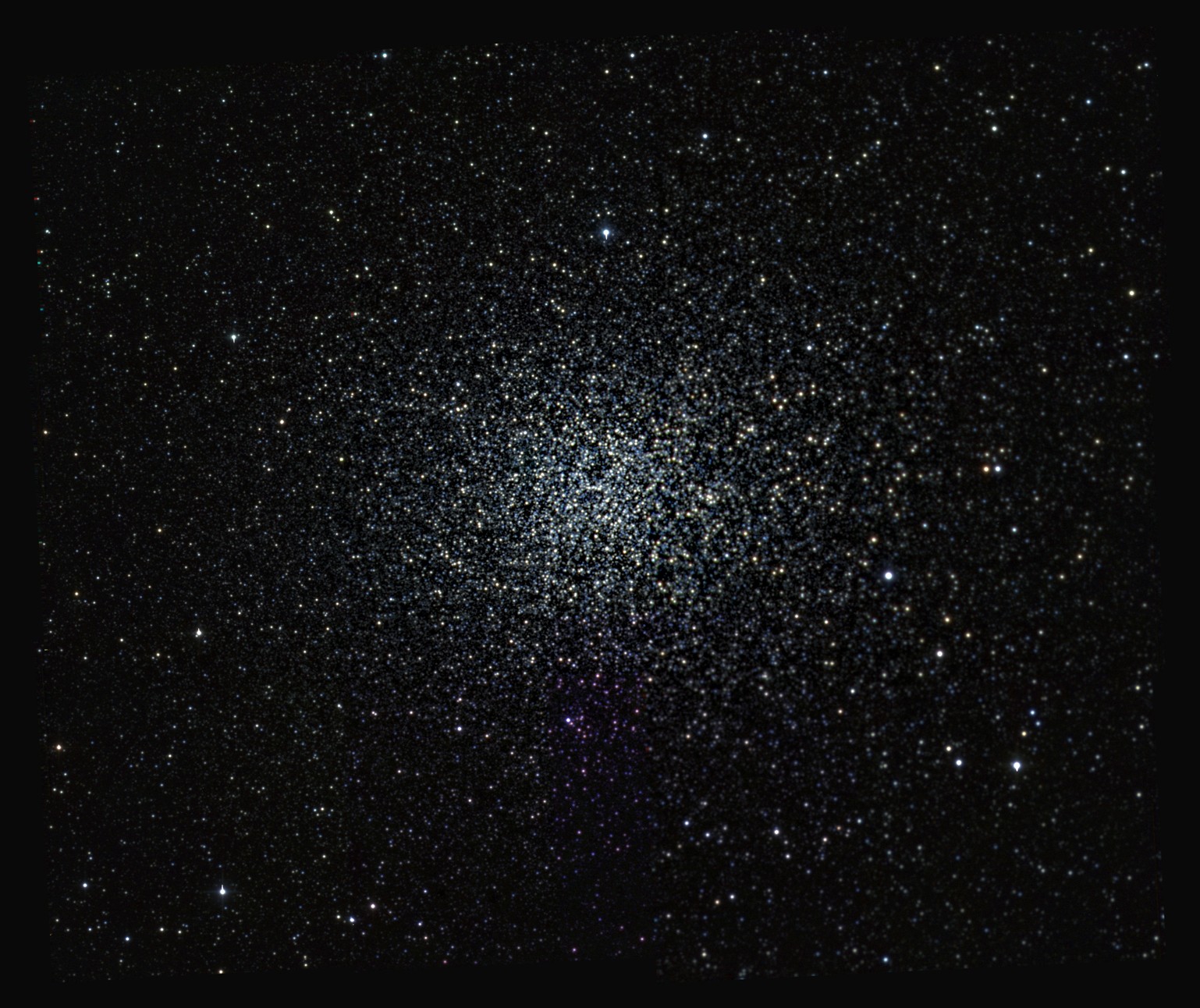}
\caption{Example LSGT images ($BVR$ composite). From top to bottom, NGC 2467 ("Skull and Crossbones nebula), NGC 2207 (an interacting galaxy pair), and NGC 5139 (Omega Centauri, a galactic globular cluster). Note that the top two images show a single frame field of view, but the bottom panel is a mosaic of 9 LSGT frames, covering a 30$^{\prime}$ by 40$^{\prime}$ field. These images were taken with one to three minute exposures.}
\end{figure}

\section{Future Upgrades}

 We hope to upgrade the LSGT system in the near future in two areas.

  First, we will work on the telescope optics to improve the image quality.
  Currently, we mostly achieve image PSF FWHM above $2\farcs0$, as shown
 in Figure 6. Better image performance has been achieved through the use of
 other PlaneWave CDK telescopes at the same site. These PlaneWave systems are very
 similar to LSGT, so there is room for improving the image quality.

  Second, we plan to install a new camera and new filters to improve the sensitivity
 and expand the wavelength coverage. We recently purchased a 1k x 1k CCD camera (iKon-M 934
 Series) from Andor, Inc. that is equipped with a back-illuminated, deep depletion CCD
 (BEX2-DD) for better sensitivity at both short and long wavelengths.
  This camera is similar to the CQUEAN camera (Park et al. 2010),
  but with an improved sensitivity at
 short wavelengths. The new camera boasts $\gtrsim 90$\% QE from 0.4 to 0.9 $\mu$m,
 with QE = 30\% at 1 $\mu$m (Figure 3). We will match it with an 18 slot filter wheel that
 will house a standard set of $grizY$ filters and a suite of medium-band filters
 with wavelength widths of 50nm for detailed studies of the spectral energy distribution
 of objects such as quasars, GRBs, AGNs, and SNe,
 and photometric reverberation mapping of AGNs. The expected date of
 operation of the new camera is in the second half of 2015.

\section{Summary}

  In this paper, we presented the characteristics and the current performance of LSGT, and our plan for future upgrades. LSGT is a 0.43-meter telescope installed at SSO, Australia that can be operated remotely from the SNU campus in Seoul, Korea.
 It is currently fully operational, with LSGT activities
 including the use for observational classes at SNU and a long term research project
 of monitoring nearby galaxies for understanding the nature of transients.
 The telescope can reach a limiting
 magnitude of $R \sim 21.5$ mag at 5$-\sigma$ with 15 min exposure
 under good weather (photometric condition), good seeing (FWHM$\sim 2\farcs2$),
 and dark sky (new moon).
   In the near future, we plan to improve the optical performance of the telescope and also
  install a new camera that will be several times more sensitive than the current camera at
  both short and long wavelengths.


\acknowledgments
 This work was supported by the Creative Initiative program, No. 2008-0060544, of the National  Research Foundation of Korea (NRFK) funded by the Korean government (MSIP).
 MI acknowledges the hospitality and the support of the Korea Institute for Advanced
 Study, where part of this work was carried out.
 This paper includes the data taken at the Siding Spring Observatory in Australia.
  We thank the staff of iTelescope.Net and SSO,
  especially Brad Moore, Peter Lake, and Lars Hansen
 for their professional assistance during the installation, the performance optimization,
 and the operation of the telescope. We thank Richard Hendick of PlaneWave Instruments for
 providing us with the telescope efficiency curve.
  We also thank Sang Gak Lee for her generous contribution to make the telescope possible.


\end{document}